\documentclass[pre,twocolumn,superscriptaddress]{revtex4-1}

\usepackage[margin=1in]{geometry}  % set the margins to 1in on all sides
\usepackage[dvipdfmx]{graphicx}    % to include figures
\usepackage{amsmath}               % great math stuff
\usepackage{amsfonts}              % for blackboard bold, etc
\usepackage{amsthm}                % better theorem environments
\usepackage{breqn}
\usepackage{latexsym, amssymb}
\usepackage{mathrsfs,enumerate}
\usepackage{graphics,epsfig,color}
\usepackage{xcolor}
\usepackage{wrapfig}

\usepackage{calrsfs}
\usepackage{bm}

\def\non{\nonumber}

\newcommand{\1}{\mbox{1}\hspace{-0.25em}\mbox{l}}

\makeatletter
\let\cat@comma@active\@empty
\makeatother

\begin{document}
	
\title{Anomalous fluctuations of renewal-reward processes with heavy-tailed distributions}

\author{Hiroshi Horii}
\author{Rapha\"el Lefevere}
\affiliation{Universit\'e de Paris, Laboratoire de Probabilit\'es, Statistiques et Mod\'elisation, UMR 8001, F-75205 Paris, France}
\author{Masato Itami}
\affiliation{Center for Science Adventure and Collaborative Research Advancement, Kyoto University, Kyoto 606-8502, Japan}
\author{Takahiro Nemoto}
\affiliation{Graduate School of Informatics, Kyoto University, Yoshida Hon-machi, Sakyo-ku, Kyoto, 606-8501, Japan}

\begin{abstract}
For renewal-reward processes  with a power-law decaying waiting time distribution, anomalously large probabilities are assigned to atypical values of the asymptotic processes. Previous works have reveals that this anomalous scaling causes a singularity in the corresponding large deviation function. In order to further understand this problem, we study in this article the scaling of variance in several renewal-reward processes: counting processes with two different power-law decaying waiting time distributions and a Knudsen gas (a heat conduction model). Through analytical and numerical analyses of these models, we find that the variances show an anomalous scaling when the exponent of the power law is -3. For a counting process with the power-law exponent smaller than -3, this anomalous scaling does not take place: this indicates that the processes only fluctuate around the expectation with an error that is compatible with a standard large deviation scaling. In this case, we argue that anomalous scaling appears in higher order cumulants.  Finally, many-body particles interacting through soft-core interactions with the boundary conditions employed in the Knudsen gas are studied using numerical simulations. We observe that the variance scaling becomes normal even though the power-law exponent in the boundary conditions is -3. 
\end{abstract}

\maketitle

\section{Introduction}

A renewal-reward process, a generalisation of continuous time Markov processes, is one of the simplest stochastic processes that can describe random sequences with memory effects \cite{asmussen2008applied, grimmett2020probability, feller2008introduction}. In contrast to its the Markov counterpart, in renewal-reward processes, the waiting time to move from one state to the next one can be distributed by a non-exponential function. The process can thus describe a broad spectrum of phenomena in physics \cite{barkai2020packets} and other fields, including a melt up of the stock market  \cite{bradley2003financial,glasserman2002portfolio} and a super spreader in epidemics  \cite{wong2020evidence,schneeberger2004scale}, where memory effects are known to be important.

When the waiting time distribution has a power law, the dynamics show a slow convergence to its stationary states due to its heavy tail. For example, the probability that the state of the system always stays in the initial state during the dynamics remains non-negligible in the large time limit \cite{lefevere2010hot}. This anomalous behaviour can be characterised using a large deviation principle (LDP) \cite{touchette2009large, den2008large}. LDP states that the logarithmic probability of a time-averaged quantity is proportional with the averaging time (with a negative proportional constant), except for the trivial probability where the time averaged quantity takes its expectation. In renewal reward processes with power-law waiting time distributions, this proportional constant, known as a rate function or large deviation function (LDF), can take the value $0$ not only for the expectation but also for a certain range of the values \cite{lefevere2010hot,lefevere2011large,lefevere2010macroscopic}. This indicates that these events are {\it more} likely to occur than in standard systems. We call this range of LDF taking the value $0$ {\it the affine part}.

The affine part tells us that these rare events occur more likely than usual, but does not tell us how likely they do. To solve this problem, finite-time analyses of the LDP are necessary. One such attempt could be a so-called strong LDP, where the next order corrections of the logarithmic probability from the LDP are computed \cite{joutard2013strong}. However, at present, it is not clear how this general theory can be extended to the case with the affine part. In \cite{tsirelson2013uniform}, Tsirelson studied a renewal-reward process with general waiting time distributions and derived the next order correction to the LDP. But he used a condition in which an affine part can not be present. Recently, in \cite{horii2022large}, the authors studied finite-time corrections of the moment generating function under the condition that the affine part appears (Theorem 2.1). Yet they did not succeed to translate it to the correction term of the LDP.

In this article, instead of focusing on the probability of rare events, we focus on the variance of the time-averaged quantities. The variance can tell us directly how much the averaged quantities fluctuate. If one considers an exponential function for the waiting time distribution, the variance of the time-averaged quantity decreases proportionally to the inverse of the averaging time because this corresponds to the case of a process having a short memory. This indicates that the averaged value mostly falls in the range around the expectation with an error that is proportional with the inverse square root  of the averaging time. In the presence of the affine part when heavy-tailed distributions are used for the waiting times, we identify, in this article, a condition under which this scaling of the variance changes.  This is consistent with the fact that heavy-tailed distributions introduce memory effects. 
Interestingly, not every power-law decaying distribution will result in this  scaling modification of the variance: We show that for distributions whose density decay faster than $1/t^3$, the variance keeps its normal scaling.
In that case, we expect that the scaling of higher order cumulants are affected, as discussed at the end of this article.

This article is organised as follows. Two models defined using renewal-reward processes are considered in this article: a counting process and a single particle model of heat conduction. These models are studied in Section~\ref{sec:countingprocess} (counting process) and  in Section~\ref{sec:Aparticleconfined} (heat conduction). Each section is organised with (i) a model introduction, (ii) introduction of renewal equations (a key tool to study the asymptotics of moments), (iii) analyses on the first moment, (iv) analyses on the second moment and variance, and (v) numerical studies. In Section~\ref{sec:discussion}, we discuss the scaling in higher order cumulants and how the scaling will change in more general heat-conduction systems.  In particular, we observe that when several particles are present and interact through soft-core interactions, the time-average of physical quantities recover a ``normal'' behaviour.  This seems to indicate that interactions break the strong memory effects that are present in the system of non-interacting particles.

\section{Counting process}
\label{sec:countingprocess}

\subsection{Model}
A renewal-reward process is a model to describe events that occur sequentially. For a given event, the next event occurs after a random waiting time (also called a renewal time or arrival time).  
The waiting times are independent-and-identically distributed random positive variables $(\tau_k)_{k\in\mathbb N}$ with a probability density $p$.
For this density, we consider the inverse Rayleigh distribution 
\begin{equation}
	\label{eq:inverseRaylei_countingprocess}
	p_\beta(\tau)=\frac{\beta}{\tau^3}{\rm exp}\left(-\frac{\beta}{2\tau^2}\right)\1(\tau>0),
\end{equation}
and the Pareto distribution 
\begin{equation}
	\label{eq:Pareto_countingprocess}
	p_\alpha(\tau)= \frac{\alpha-1}{(1+\tau)^\alpha}\1(\tau>0)
\end{equation}
with $\alpha=3$, both of which do not have a finite second moment, {\it i.e.}, $\mathbb E[\tau^2]=\infty$.  The main quantity of interest in this section is the number of events that have occured up to time $t>0$. This is the counting process $N_t$
\begin{equation}
\label{eq:def_countingprocess}
N_t=\sup\{k:S_k\leq t\},
\end{equation}
where $S_k=\tau_1+\ldots+\tau_k$. 
We denote its $q$-th order moment by $m_q(t)$:
\begin{equation}
m_q(t):=\mathbb E[N_t^q].
\end{equation}

Note that with respect to \cite{metzler2014anomalous}, we consider the case where the expectation of the waiting time is finite and the renewal theorem \cite{asmussen2008applied} implies that the counting process $N_t$ behaves as $N_t\sim t / {\mathbb E}[\tau]$ for $t\to\infty$.  We study the fluctuations around that behaviour.

\subsection{Renewal equations}

To analyse the asymptotics of $m_q(t)$, we rely on renewal equations: a powerful tool to analyse renewal-reward processes.
From a straightforward computation, one can establish the following renewal equation for $m_1(t)$ \cite{grimmett2020probability}:
\begin{equation}
m_1(t)=F(t)+\int_0^t\; ds\; m_1(t-s) p(s),
	\label{eq:renewalequation1}
\end{equation}
where $F$ is the cumulative waiting time distribution function. From this equation,  a simple expression for the Laplace transform of $m_1(t)$ is derived. Defining the Laplace transform of a function $f$ by
\begin{equation}
\tilde f(s)	:=\int^{\infty}_{0}e^{-st}f(t) dt,
\end{equation}
we then derive, from the equation \eqref{eq:renewalequation1}, 
\begin{eqnarray}
	\tilde{m}_1(s) 
	&=&\frac{\tilde{F}(s)}{1-s\tilde{F}(s)},
	\label{basicr}
\end{eqnarray}
where we have used $\tilde{p}(s)=s\tilde{F}(s)$.

Similarly, one can also derive a renewal equation for $m_2(t)$,
\begin{equation}
\begin{split}
	m_2(t)  = & \int_0^t\mathbb E[N_{t-s}^2]p(s) ds \\ 
	&+2 \int_0^t m_1(t-s) p(s)\,ds+F(t),
\label{renewc2}
\end{split}
\end{equation}
from which  the Laplace transform of $m_2(t)$ is obtained as
\begin{equation}
	\tilde m_2(s)=\tilde m_1(s)(1+2s\tilde m_1(s)).
	\label{c2c1}
\end{equation}

Moreover, a renewal equation for the moment-generating function can be derived. (See Appendix~\ref{sec:appendix_kthmoment}). From the equation, we derive the Laplace transform of $m_q(t)$ as
\begin{align}
 \tilde{m}_{q}(s) = \sum_{k=1}^{q} \left[ \sum_{i=1}^{k}\begin{pmatrix}k\\i\end{pmatrix}i^{q}(-1)^{k-i}\right] s^{k-1}\left[ \tilde{m}_{1}(s)\right]^{k}.
\end{align}

\subsection{Convergence of the first moment}

When a waiting-time density $p$ has a finite mean $\mathbb E[\tau]=\mu$ and a finite variance $\sigma^2$, Feller has proven (Chapter 11, section3, theorem 1) \cite{feller2008introduction} that
\begin{equation}
\frac{m_1(t)}{t}-\frac{1}{\mu}\sim \frac{\sigma^2-\mu^2}{2\mu^2 t}.
\label{conv_feller} 	
\end{equation}
This result can be easily derived by using the following expansion: 
\begin{equation}
 s \tilde F(s)=1-\mu s+ (\sigma^2+\mu^2) \frac{s^2}{2}+o(s^2).
\label{mgf}	
\end{equation}
Indeed, by inserting it into (\ref{basicr}), we get
\begin{equation}
\tilde m_1(s)= \frac{1}{\mu s^2}+\frac{\sigma^2-\mu^2}{2\mu^2 s}+o\left(\frac{1}{s}\right), 
\label{m1laplace}	
\end{equation}
which leads to
\begin{equation}
m_1(t)= \frac 1 \mu t+ \frac{\sigma^2-\mu^2}{2\mu^2 }+o(1).
\label{eq:tauberian_for_m}
\end{equation}
A rigolous justification to derive \eqref{eq:tauberian_for_m} from \eqref{m1laplace} is based on the Tauberian theorem~\cite{feller2008introduction}. See Appendix~\ref{sec:tauberian} for more details. From this argument, we can see that the condition $\mathbb E[\tau^2]=\infty$ is necessary for $m_1(t)$ to have an anomalous scaling.  For this reason, we study in this section the two waiting-time distributions behaving at infinity like $1/t^3$.

Let us first consider the case of the inverse Rayleigh distribution. 
Let 
\begin{equation}
	\phi(s)=\int_0^\infty e^{-st}e^{-\frac {1}{2t^2}}\,dt.
	\label{phi}
\end{equation}
We then have for its cumulative distribution function,
\begin{equation}
	\tilde F_\beta(s)=\beta^{\frac 1 2}\phi(\beta^{\frac 1 2} s),
\end{equation}
and 
\begin{equation}
	\label{eq:laplacem}
	\tilde{m}_1(s)=\frac{\beta^{\frac 1 2}\phi(\beta^{\frac 1 2} s)
}{1-s \beta^{\frac 1 2}\phi(\beta^{\frac 1 2} s),
}
\end{equation}
from \eqref{basicr}.
We then expand $\phi(s)$ in $s$:
\begin{equation}
\phi(s)=\frac 1 s-\sqrt{\frac{\pi}{2}}-\frac 1 2 s\ln (s)+O(s),
	\label{expandphi}
\end{equation}
leading to
\begin{equation}
	\label{eq:expandm}
\tilde m_1(s)=\sqrt{\frac{2}{\beta\pi}}\frac 1 {s^2}-\frac{1}{\pi s}\ln(s)+o\left(\frac{\ln (s)}{s}\right).
\end{equation}
By using the Tauberian theorem (Appendix~\ref{sec:tauberian}), we obtain 
\begin{equation}
	\label{eq:Rayleighm1}
	\frac{m_1(t)}{t} - \sqrt{\frac{2}{\beta\pi}}  = \frac{\ln(t)}{ t \pi} + o\left ( \frac{\ln(t)}{t} \right ),
\end{equation}
for large $t$. This is to be compared to (\ref{conv_feller}): we see that the convergence is slower in our case.

We can repeat the same analysis in the case of the Pareto distribution. The cumulative distribution is derived as
\begin{eqnarray}
	\non\\
	F_3(t):=\mathbb P[\tau\leq t]&=&\left\{\begin{array}{ll}0 & t\leq 0\\ 
		1-\frac{1}{(1+t)^{2}}& t >0.
	\end{array}
\right.
\end{eqnarray}
when $m=3$. We insert the Laplace transform of $F_3$ in (\ref{basicr}) and again look at the expansion around $s$ of $\tilde m$ and get 
\begin{equation}
	\label{eq:solutionmsp}
	\tilde{m}_1(s)= \frac{1}{s^2}-\frac{\ln(s)}{s}+o\left(\frac{\ln (s)}{s}\right).
\end{equation}
We thus obtain the following behaviour for $m(t)$ for large $t$:
\begin{equation}
	\label{eq:convergenceNtp}
	\frac{m_1(t)}{t} - 1 =  \frac{{\rm ln}(t)}{t} + o\left ( \frac{\ln(t)}{t} \right ).
\end{equation}

\subsection{Convergence of the variance}
We then study the large time behaviour of the variance
\begin{equation}
c_2(t)= \frac{m_2(t) - m_1(t)^2}{t^2}.
\label{defc2}
\end{equation}
In the case that a waiting time density $p$ has 
a finite mean $\mathbb E[\tau]=\mu$ and a finite variance  $\sigma^2$, we obtain from \eqref{c2c1} and (\ref{m1laplace})
\begin{equation}
	\tilde m_2(s)
	=\frac{2}{\mu^2 s^3}+\frac{1}{s^2}\frac{1}{\mu}\left (\frac{2\sigma^2-\mu^2}{\mu^2} \right )+o\left (\frac{1}{s^2} \right ),
\end{equation}
which yields
\begin{equation}
m_2(t)=\frac{1}{\mu^2}t^2+\frac{1}{\mu}\left(\frac{2\sigma^2-\mu^2}{\mu^2}\right)t+o(t)
\end{equation}
with the aid of the Tauberian theorem (\ref{tauberian}). $c_2(t)$ is finally obtained as
\begin{equation}
c_2(t)=\frac{\sigma^2}{\mu^3t}+o\left(\frac 1 t\right).
\end{equation}

Let us now consider the case of the inverse Rayleigh distribution. Inserting the expression (\ref{eq:expandm}) for $\tilde m_1(s)$ in (\ref{c2c1}), we obtain,
\begin{equation}
\tilde m_2(s)= \frac{4}{\beta\pi}\frac{1}{s^3}-\frac{4\sqrt{2}}{\sqrt{\beta}\pi^{3/2}}\frac{\ln s}{s^2}+o\left(\frac{\ln(s)}{s^2}\right),
\end{equation}
and then
\begin{equation}
m_2(t)=\frac{2}{\beta\pi}t^2+\frac{4\sqrt{2}}{\sqrt{\beta}\pi^{3/2}}t\ln(t)+o(t\ln (t)).
\end{equation}
Therefore
\begin{equation}
	\label{eq:RayleighNtvar}
	c_2(t)=\frac{2\sqrt{2}}{\sqrt{\beta}\pi^{3/2}} \frac{\ln(t)}{t}+o\left(\frac{\ln(t)}{t}\right).
\end{equation}
Proceeding in the same way for the Pareto distribution, we obtain in that case 
\begin{equation}
	\label{eq:ParetoNtvar}
	c_2(t)=2 \frac{\ln t}{t}+o\left(\frac{\ln (t)}{t}\right).
\end{equation}

\subsection{Numerical study}
\begin{figure*}[htbp]
	\begin{center}
		\includegraphics[clip,width=15.cm]{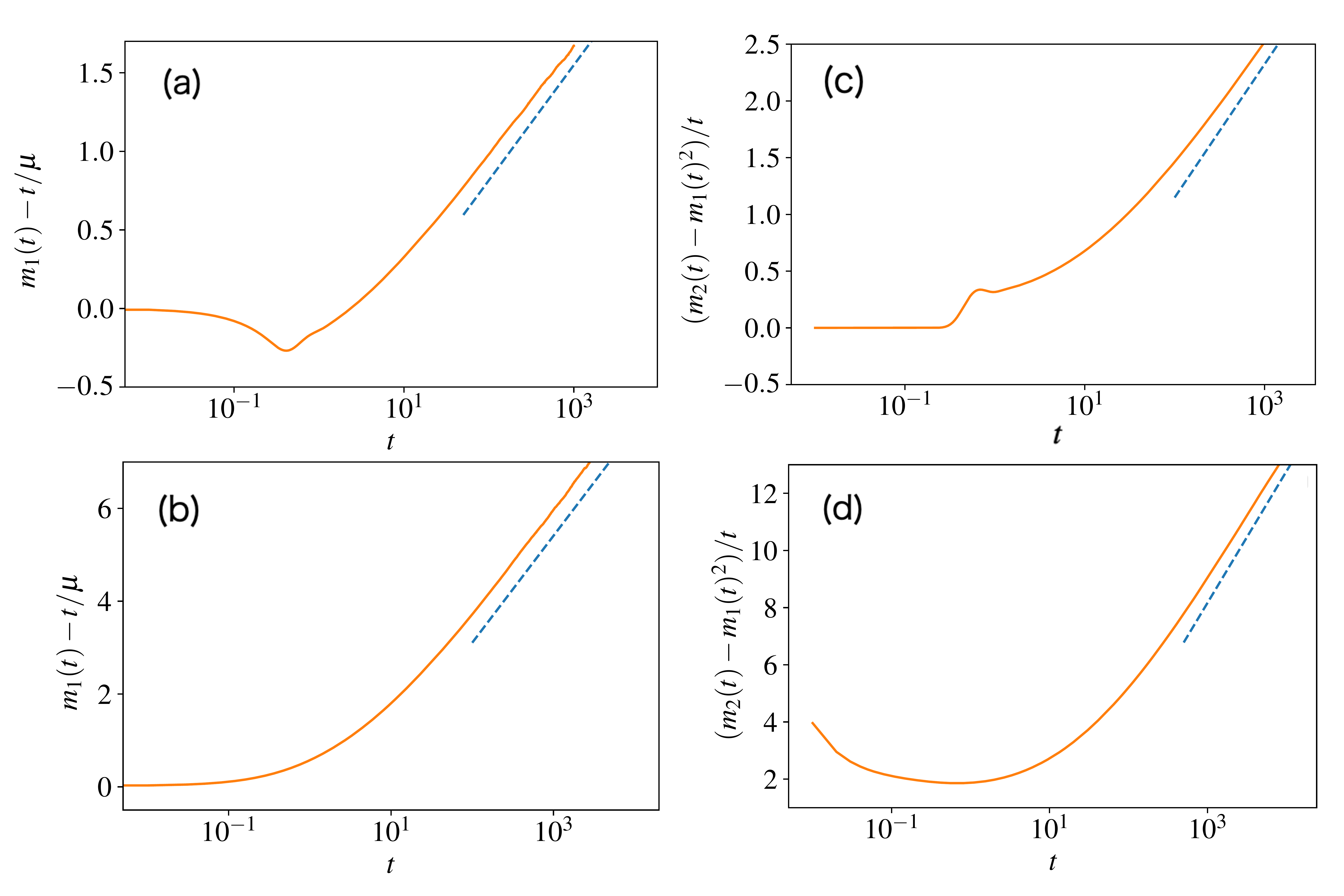} 
		\caption{{\bf (a,b)} $m_1(t) - t /\mu$  obtained from numerical simulations of the counting process $N_t$ (with $10^8$ samples) are plotted  as a function of time in log-scale as orange lines.  		
		 For the inverse Rayleigh waiting time distribution (a), $\beta=1$ and $\mu=1/\sqrt{2/(\beta \pi)}$, while for the Pareto waiting time distribution (b), $m=3$ and $\mu=1$. 		 
		 $\ln(t) / \pi + \rm const.$ and $ \ln(t) + \rm const. $ are also plotted as blue dashed lines for (a) and (b). 
		 {\bf (c,d)}  $(m_2(t)-m_1(t)^2)/t$ obtained from the same numerical simulations are plotted as a function of time as orange lines for the inverse Rayleigh waiting time distribution (c) and for the Pareto waiting time distribution (d). $\frac{2\sqrt{2}}{\sqrt{\beta}\pi^{3/2}} \ln(t) + \rm const.$ for (c) and $2 \ln (t) + \rm const.$ for (d) are also plotted as blue dashed lines in the same figures.  
	         The agreements between the slopes of orange lines and those of blue lines in these semi-log graphs demonstrate the validity of \eqref{eq:Rayleighm1}, \eqref{eq:convergenceNtp}, \eqref{eq:RayleighNtvar} and \eqref{eq:ParetoNtvar}, as detailed in the main text. }
		\label{fig:numerics_Nt}
	\end{center}
\end{figure*}

We perform  numerical simulations of  the counting process $N_t$ to illustrate the accuracy of \eqref{eq:Rayleighm1}, \eqref{eq:convergenceNtp},  \eqref{eq:RayleighNtvar} and \eqref{eq:ParetoNtvar}. First, $m_1(t)-t \sqrt{2/(\beta \pi)}$ (resp. $m_1(t)-t$) computed from the numerical simulations is plotted as an orange line in Fig.\ref{fig:numerics_Nt}(a) (resp. Fig.\ref{fig:numerics_Nt}(b)) for the inverse Rayleigh (resp. Pareto) waiting time distribution. 
According to \eqref{eq:Rayleighm1} and \eqref{eq:convergenceNtp}, these lines are equivalent to
$\ln(t)/\pi + o(\ln(t))$ and $\ln(t) + o(\ln(t))$. Assuming that these $o(\ln(t))$ terms are constant over time when $t$ is large, we next plot $\ln (t)/\pi + \rm const.$ (Fig.\ref{fig:numerics_Nt}(a)) and $\ln (t)+ \rm const.$ (Fig.\ref{fig:numerics_Nt}(b)) in the same figures.

We then plot $(m_2(t)-m_1(t)^2)/t$ computed from the same numerical simulations in Fig.\ref{fig:numerics_Nt} (c,d) for the inverse Rayleigh (Fig.\ref{fig:numerics_Nt}(c))
 and the Pareto (Fig.\ref{fig:numerics_Nt}(d)) waiting time distributions. Reference lines  $\frac{2\sqrt{2}}{\sqrt{\beta}\pi^{3/2}} \ln(t) +  \rm const.$  (Fig.\ref{fig:numerics_Nt}(c))  and $2 \ln (t)+ \rm const.$ (Fig.\ref{fig:numerics_Nt}(d)) are also plotted in the same figures. In these four figures, we observe good agreements between the slopes of the reference lines and the results of numerical simulations in semi-log scale. This demonstrates the validity of  \eqref{eq:Rayleighm1}, \eqref{eq:convergenceNtp},  \eqref{eq:RayleighNtvar} and \eqref{eq:ParetoNtvar}.

\section{A particle confined between two hot walls}
\label{sec:Aparticleconfined}

Our aim in this section is to show that the slow convergence of the renewal function of processes having density $\sim 1/t^3$ as $t\to\infty$ also holds for physical observables in a Knudsen gas \cite{lebowitz1957model}. 
For this, let us consider the model of a single particle bouncing back between two thermal walls. 

\subsection{Model} \label{subsec:Model}

We consider a  particle in a one-dimensional box that has two different temperatures at both ends. The confined tracer moves freely in the box of size $1$ and is reflected at the end of the box with a random speed $v$ distributed according to the following Rayleigh distribution:
\begin{equation}
	\label{eq:Rayleigh}
	q_{\beta_{\pm}}(v) = \beta_{\pm} v e^{-\beta_{\pm}\frac{v^2}{2}}\1(v>0),
\end{equation}
where $\beta_{+}=1/T_{+}$ (resp. $\beta_{-}=1/T_{-}$) is the inverse temperature of the right (resp. left) wall. 
\begin{figure}[htbp]
	\begin{center}
		\includegraphics[clip,width=7.0cm]{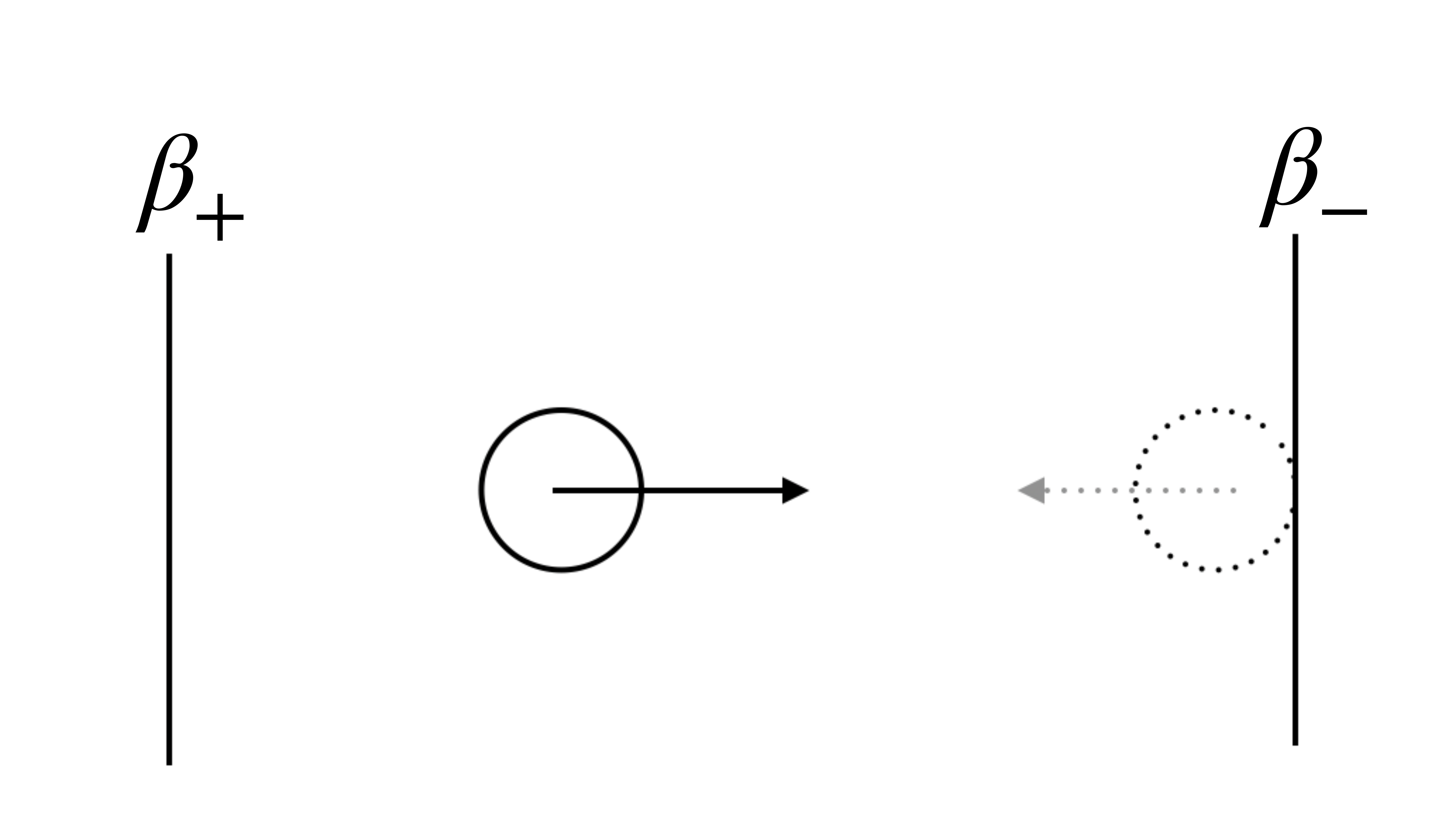}
		\caption{Schematic figure to explain the setup of the 1 particle model. When the particle moves to the right (resp. left) wall, $\sigma_k=1$ (resp. $-1$)}
		\label{fig:model1}
	\end{center}
\end{figure}

Let $x_0\in [0,1]$ and $v_0$ the initial position and velocity of the particle and $\sigma_0=v_0/|v_0|$.
We denote the initial condition by $\theta$, {\it i.e., $\theta=(x_0,v_0)$}.
The first time that the particle hits a wall is given by $S_{\theta,0} = (\frac{1}{2}(\sigma_0+1)-x_0)/v_0$, and the subsequent hitting times are given by
\begin{equation}
S_{\theta,k}=S_{\theta,0}+1/v_1+\ldots+1/v_k, \quad k\geq 1,
\end{equation}
where $v_k$ is a random variable distributed according to a law $q_{\beta_{\sigma_k}}$ and $\sigma_k=(-1)^k\sigma_0$.
This may be rewritten as
\begin{equation}
S_{\theta,k}=S_{\theta,0}+\tau_1+\ldots+\tau_k, \quad k\geq 1
\end{equation}
with the sequence of independent waiting times $(\tau_k)_{k\in\mathbb N}$ distributed with the inverse Rayleigh distribution
$p_{\beta_k}(\tau)$ defined as \eqref{eq:inverseRaylei_countingprocess}. 
The energy exchanged between the two walls during a time interval $[0,t]$ is defined as
\begin{equation}
	\label{eq:current}
	J_\theta (t):= \frac{1}{2}\sum_{k = 1}^{N_t}v_k^2\sigma_k,
\end{equation}
if $t\geq S_0^\theta$ and $J_\theta(t)=0$ otherwise, where $N_t$ is the counting process \eqref{eq:def_countingprocess}, 
We denote by $m_{\theta,q} (t)$ the $q$-th moment of $J_\theta (t)$:
\begin{eqnarray}
	m_{\theta,q}(t)= \mathbb{E}[J^q_\theta(t)].
\end{eqnarray}

A generalisation to the system with an arbitrary box size $L$ is straightforward. Indeed, denoting by $S^{L}_{\theta,k}$ the corresponding hitting times with the boundaries, it is easy to see that $S^{L}_{\theta,k}=L S^{1}_{\theta,k}$. This indicates  that $N^L_t=N^1_{t/L}$ where $N^L_t$ denotes the counting process corresponding to the hitting times $S^{L}_{\theta,k}$. 
For the energy current in a box of size $L$, we also have that 
\begin{equation}
	J_{\theta}^{L}(t)=J_{\theta}^{1}\left(\frac{t}{L}\right).
	\label{scalingJ}
\end{equation}
In the following we perform all computations with the case $L=1$ and then obtain the result for an arbitrary $L>0$ by using this scaling relation.

\subsection{Convergence of the current: first moment}
\label{sec:convergence_current_first}

For simplicity, we consider only the following two types of initial conditions:
\begin{equation}
\theta_{+} = (0,v_0)
\end{equation}
with $v_0 < 0$ and
\begin{equation}
\theta_{-} = (1,v_0)
\end{equation}
with $v_0 > 0$, {\it i.e.}, the cases of a particle just before hitting the left wall (temperature $\beta_+$) and of a particle just before hitting the right wall (inverse temperature $\beta_-$).
As the particle immediately hits each wall when the process starts, the value of the initial velocity $v_0$ is unimportant. We thus denote by $+$ the initial condition $\theta_+$ and by $-$ the initial condition $\theta_-$.

Dynamics with these two initial conditions are related via the renewal property:
\begin{equation}
	\mathbb{E}[{J_\pm(t)}\mid \tau_1=u]=\frac{\pm 1}{2u^2}+\mathbb{E}[{J_{\mp}(t-u)}] ,
	\label{renewJ}
\end{equation}
if $0\leq u\leq t$ and $\mathbb{E}[{J_{\pm}(t)}\mid \tau_1=u]=0$ if $u>t$.
This means that the process conditioned by the first-waiting time (the left-hand side) is equal to the other process with some increments (the right-hand side). 
By integrating (\ref{renewJ}) with respect to the inverse Rayleigh waiting time density \eqref{eq:inverseRaylei_countingprocess},
we obtain the following coupled renewal-reward equations for the currents
\begin{align}
 m_{-,1}(t) &= -\left(\frac{1}{2t^2}+\frac{1}{\beta_{-}} \right)e^{-\frac{\beta_{{-}}}{2t^2}}
 \notag\\
 & \qquad + \int_0^{t} du\; m_{+,1}(t-u)p_{\beta_-}(u), \label{eq:REr}
 \\
 m_{+,1}(t) &= +\left(\frac{1}{2t^2}+\frac{1}{\beta_{+}} \right)e^{-\frac{\beta_{{+}}}{2t^2}}
 \notag\\
 & \qquad + \int_0^{t} du\; m_{-,1}(t-u)p_{\beta_+}(u). \label{eq:REl}
\end{align}
In order to derive the speed of convergence of the current, we perform a Laplace transform of \eqref{eq:REr} and \eqref{eq:REl}:
\begin{equation}
	\label{eq:RErf}
	\tilde{m}_{-,1}(s)=-\tilde{H}_-(s)-\frac{1}{\beta_-}\tilde{F}_{\beta_-}(s)+s\,\tilde{m}_{+,1}(s)\tilde{F}_{\beta_-}(s),
\end{equation}
\begin{equation}
	\label{eq:RElf}
	\tilde{m}_{+,1}(s)=\tilde{H}_+(s)+\frac{1}{\beta_+}\tilde{F}_{\beta_+}(s)+s\,\tilde{m}_{-,1}(s)\tilde{F}_{\beta_+}(s),
\end{equation}
where $\tilde{H}_{\pm}(s)$ is the Laplace transform of $H_{\pm}(t)=\frac{1}{2t^2}e^{-\frac{\beta_{\pm}}{2t^2}}$ and $\tilde{F}_{\beta_{\pm}}(s)$ is the Laplace transform of the cumulative inverse Rayleigh distribution. By substituting \eqref{eq:RErf} into \eqref{eq:RElf}, we then obtain an equation for $\tilde{m}_{+,1}(s)$ as
\begin{align}
 \label{eq:laplacecurrent1}
 \tilde{m}_{+,1}(s)&=\frac{\tilde{H}_+(s)+\frac{1}{\beta_+}\tilde{F}_{\beta_+}(s)}{1-s^2\tilde{F}_{\beta_+}(s)\tilde{F}_{\beta_-}(s)}
 \notag\\
 &\qquad - s\tilde{F}_{\beta_+}(s)\frac{\tilde{H}_-(s)+\frac{1}{\beta_-}\tilde{F}_{\beta_-}(s)}{1-s^2\tilde{F}_{\beta_+}(s)\tilde{F}_{\beta_-}(s)},
\end{align}
which leads to
\begin{dmath}
	\label{eq:laplacej}
	\tilde{m}_{\pm,1}(s)= \kappa\left(\frac{1}{\beta_+}-\frac{1}{\beta_-}\right)\frac{1}{s^2} -\kappa^2 \frac{(\beta_+ + \beta_-)}{2} \left(\frac{1}{\beta_+}-\frac{1}{\beta_-}\right)\frac{{\rm ln}(s)}{s} + o\left(\frac{\ln (s)}{s}\right),
\end{dmath}
where $\kappa$ is the conductivity given by
\begin{equation}
	\kappa^{-1}=\left(\frac{\pi\beta_-}{2} \right)^{\frac{1}{2}}+\left(\frac{\pi\beta_+}{2} \right)^{\frac{1}{2}}.
\end{equation}
Using again the Tauberian theorem for Laplace transform (Appendix~\ref{sec:tauberian}), we finally get,
\begin{dmath}
	\label{eq:mj}
	\frac{m_{\pm,1}(t)}{t}= \kappa\left(\frac{1}{\beta_+}-\frac{1}{\beta_-}\right)+\kappa^2 \frac{(\beta_+ + \beta_-)}{2} \left(\frac{1}{\beta_+}-\frac{1}{\beta_-}\right) \frac{\ln t}{t}+o\left(\frac{\ln (t)}{t}\right).
\end{dmath}
Note that the asymptotic form of the average current $m_{+,1}(t)$ and $m_{-,1}(t)$ have opposite signs, but this is because the definition of the current includes $(-1)^{\pm 1}$ term: these two expressions are  physical equivalent. For the average current in a box of size $L$, we get
\begin{dmath}
	\label{eq:mjL}
\frac{m^L_{\pm,1}(t)}{t}=\frac{1}{L}\frac{m^1_{\pm,1}(\frac{t}{L})}{\frac{t}{L}}= \frac{\kappa}{L}\left(\frac{1}{\beta_+}-\frac{1}{\beta_-}\right)+\kappa^2 \frac{(\beta_+ + \beta_-)}{2} \left(\frac{1}{\beta_+}-\frac{1}{\beta_-}\right) \frac{\ln (t)}{t}+o\left(\frac{\ln (t)}{t}\right).	
\end{dmath}

\subsection{Variance of the current of energy  between heat baths}
We next discuss the large time asymptotics of the variance of the current. 
The renewal property for the second moment of the current is expressed by 
\begin{dmath}
	\mathbb{E}[{J^2_{\pm}(t)}\mid \tau_1=u]=\left[\frac{1}{2u^2}\right]^2+\frac{\sigma}{u^2}\mathbb{E}[{J_{\mp}(t-u)}] +\mathbb{E}[{J^2_{\mp}(t-u)}]
	\label{renewJ2}
\end{dmath}
 if $0\leq u\leq t$ and $\mathbb{E}[{J_{\pm}(t)}\mid \tau_1=u]=0$ if $u>t$.  Let us introduce for $t>0$,  
\begin{eqnarray}
	L_{\pm}(t)&=&\left(\frac{1}{4t^4}+\frac{1}{\beta_{\pm}t^2}+\frac{2}{\beta^2_{\pm}} \right)e^{-\frac{\beta_{{\pm}}}{2t^2}},\non\\
	g_{\pm}(t)&=&\frac{\beta_{\pm}}{t^5}e^{-\frac{\beta_{{\pm}}}{2t^2}}\non.
\end{eqnarray}
Then, integrating \eqref{renewJ2} 
with respect to the inverse Rayleigh waiting time density \eqref{eq:inverseRaylei_countingprocess}
and using the relation
\begin{equation}
	\int_0^{t}du\,\frac{1}{4u^4} p_{\beta_{\pm}}(u)
	= \left(\frac{1}{4t^4}+\frac{1}{\beta_{\pm}t^2}+\frac{2}{\beta^2_{\pm}} \right)e^{-\frac{\beta_{{\pm}}}{2t^2}}
\end{equation}
 for $t>0$, the renewal equations for the second moment of the current  are derived as
\begin{align}
 m_{+,2}(t)&= L_+(t)+\int_0^{t} du\, m_{-,1}(t-u)g_+(u)
 \notag\\
 & \qquad + \int_0^{t} du\, m_{-,2}(t-u)p_{\beta_+}(u) ,
 \label{eq:REr2}\\
 m_{-,2}(t)&=  L_-(t)-\int_0^{t} du\, m_{+,1}(t-u)g_-(u)
 \notag\\
 & \qquad + \int_0^{t} du\, m_{+,2}(t-u)p_{\beta_-}(u),
 \label{eq:REl2}
\end{align}
In order to derive the large time asymptotic of the second moment of the current, we perform Laplace transform of \eqref{eq:REr2} and \eqref{eq:REl2}, 
\begin{equation}
	\label{eq:RErf2}
	\tilde{m}_{+,2}(s)=\tilde{L}_+(s)+\tilde{m}_{-,1}(s)\tilde{g}_+(s)+s\,\tilde{m}_{-,2}(s)\tilde{F}_{\beta_+}(s),
\end{equation}
\begin{equation}
	\label{eq:RElf2}
	\tilde{m}_{-,2}(s)=\tilde{L}_-(s)-\tilde{m}_{+,1}(s)\tilde{g}_-(s)+s\,\tilde{m}_{+,2}(s)\tilde{F}_{\beta_-}(s).
\end{equation}
We solve these linear equations for $\tilde{m}_{-,2}(s)$ and $\tilde{m}_{+,2}(s))$ by using
$\tilde m_{\pm,1}(s)$ obtained in section~\ref{sec:convergence_current_first} and the following expansions of $\tilde L_\pm(s)$ and $\tilde g_\pm(s)$
\begin{equation}
	\tilde L_\pm(s)=\frac{2}{\beta^2_\pm s}-\frac{3}{4\beta^{3/2}_\pm}\sqrt{\frac{\pi}{2}}+\frac{s}{4\beta_\pm}+O(s^2),
\end{equation}
\begin{equation}
	\tilde g_\pm(s)=\frac{2}{\beta_\pm}-\frac{1}{\sqrt{\beta_\pm}}\sqrt{\frac{\pi}{2}} s+O(s^2).
\end{equation}
Recalling
$\tilde F_{\beta_\pm}(s)=\beta_{\pm}^{\frac 1 2}\phi(\beta_{\pm}^{\frac 1 2} s)$ with
\begin{equation}
\phi(s)=\frac 1 s -\sqrt{\frac{\pi}{2}}-\frac{1}{2}s\ln (s)+O(s),
\end{equation}
the Laplace transform of the second moment of the current is  derived as
\begin{dmath}
	\tilde{m}_{\pm,2}(s)=2\kappa^2\left(\frac{1}{\beta_+}-\frac{1}{\beta_-}\right)^2\frac{1}{s^3}-\frac{2\kappa^3\ln(s)}{s^2}(\beta_++\beta_-)\left(\frac{1}{\beta_+}-\frac{1}{\beta_-} \right)^2+o\left(\frac{\ln(s)}{s^2}\right).
\end{dmath}
From the Tauberian theorem (Appendix~\ref{sec:tauberian}), we finally arrive at the asymptotic form of the variance 
\begin{eqnarray}
	\label{eq:convergencevarcurrent}
	{\mathrm {Var}}\left(\frac{J_\pm(t)}{t}\right)&=& \frac{m_{\pm,2}(t)}{t^2}-\frac{(m_{\pm,1}(t))^2}{t^2}\nonumber\\
	&=&	 \kappa^3(\beta_++\beta_-)\left(\frac{1}{\beta_+}-\frac{1}{\beta_-} \right)^2\frac{\ln(t)}{t}\non\\
	&&+o\left(\frac{\ln (t)}{t}\right).
\end{eqnarray}
 This result agree with our previous work \cite{lefevere2011large}. As for the variance of the current in a box of size $L$, we get:
\begin{align}
 \label{eq:convergencevarcurrentL}
 {\mathrm {Var}}\left(\frac{J^L_{\pm}(t)}{t}\right)&=\frac{1}{L^2}{\mathrm {Var}}\left(\frac{J^1_\pm(\frac{t}{L})}{t/L}\right)
 \notag\\
 &= \frac{1}{L} \kappa^3(\beta_++\beta_-)\left(\frac{1}{\beta_+}-\frac{1}{\beta_-} \right)^2\frac{\ln(t)}{t}
 \notag\\
 &\qquad +o\left(\frac{\ln (t)}{t}\right).
\end{align}

\subsection{Convergence of the thermal energy}

A similar formulation can be applied to study the convergence of the  time-averaged  kinetic energy  defined as
\begin{eqnarray}
	&&E_\pm (t)=\frac{1}{2}\sum_{k=1}^{N_t}\frac{1}{\tau_k}.\non\\
	\end{eqnarray}
The expected values of the energy are denoted by
\begin{eqnarray}
	m_E^{\pm}(t)= \mathbb{E}[E_\pm(t)].
\end{eqnarray}
As in the section~\ref{sec:convergence_current_first}, we construct two equations in Laplace space 
\begin{equation}
	\tilde{m}_E^+(s)=\tilde{h}_+(s)+\tilde{g}_+(s)+s\tilde{m}_E^-(s)\tilde{F}_{\beta_+}(s),
\end{equation}
\begin{equation}
	\tilde{m}_E^-(s)=\tilde{h}_-(s)+\tilde{g}_-(s)+s\tilde{m}_E^+(s)\tilde{F}_{\beta_-}(s),
\end{equation}
Here, $h(t)$ and $g(t)$ are given by
\begin{eqnarray}
h_\pm(t)&=& \frac{e^{-\frac{\beta_{\pm}}{2t^2}}}{2t},\\
g_\pm(t)&=&\frac{1}{2}\sqrt{\frac{\pi}{2\beta_{\pm}}} {\rm Erfc}\left[\sqrt{\frac{\beta_{\pm}}{2}}\frac{1}{t}\right].
\end{eqnarray}
Thus,  $\tilde{m}_E^+(s)$ is calculated as
\begin{align}
 \label{eq:laplaceme}
 \tilde{m}_E^+(s)&=\frac{\tilde{h}_+(s)+\tilde{g}_+(s)}{1-s^2\tilde{F}_{\beta_+}(s)\tilde{F}_{\beta_-}(s)}
 \notag\\
 & \qquad +s\tilde{F}_{\beta_+}(s)\frac{\tilde{h}_-(s)+\tilde{g}_-(s)}{1-s^2\tilde{F}_{\beta_+}(s)\tilde{F}_{\beta_-}(s)}.
\end{align}
Proceeding in tha same way as for the current, we can expand the functions involved for small $s$ and obtain
\begin{dmath}
 \tilde{m}_E^+(s)= \sqrt{\frac{\pi}{8}}\kappa\left(\sqrt{\frac{1}{\beta_+}}+\sqrt{\frac{1}{\beta_-}}\right)\frac{1}{s^2} -\frac{1}{4}\sqrt{\frac{\pi}{2}}\kappa^2 (\beta_+ + \beta_-) \left(\sqrt{\frac{1}{\beta_+}}+\sqrt{\frac{1}{\beta_-}}\right)\frac{{\rm ln}(s)}{s} + o\left(\frac{\ln (s)}{s}\right).
\end{dmath}
As with the derivation of \eqref{eq:mj}, the large time asymptotics of $m_E^{\pm}(t)$ are derived as
\begin{dmath}
	m_E(t)= \sqrt{\frac{\pi}{8}}\kappa\left(\sqrt{\frac{1}{\beta_+}}+\sqrt{\frac{1}{\beta_-}}\right)t+\frac{1}{4}\sqrt{\frac{\pi}{2}}\kappa^2 (\beta_+ + \beta_-) \left(\sqrt{\frac{1}{\beta_+}}+\sqrt{\frac{1}{\beta_-}}\right)\ln (t)+o(\ln (t)).
\end{dmath}

\subsection{Numerical simulations}

\begin{figure}[htbp]
	\begin{center}
		\includegraphics[clip,width=7.5cm]{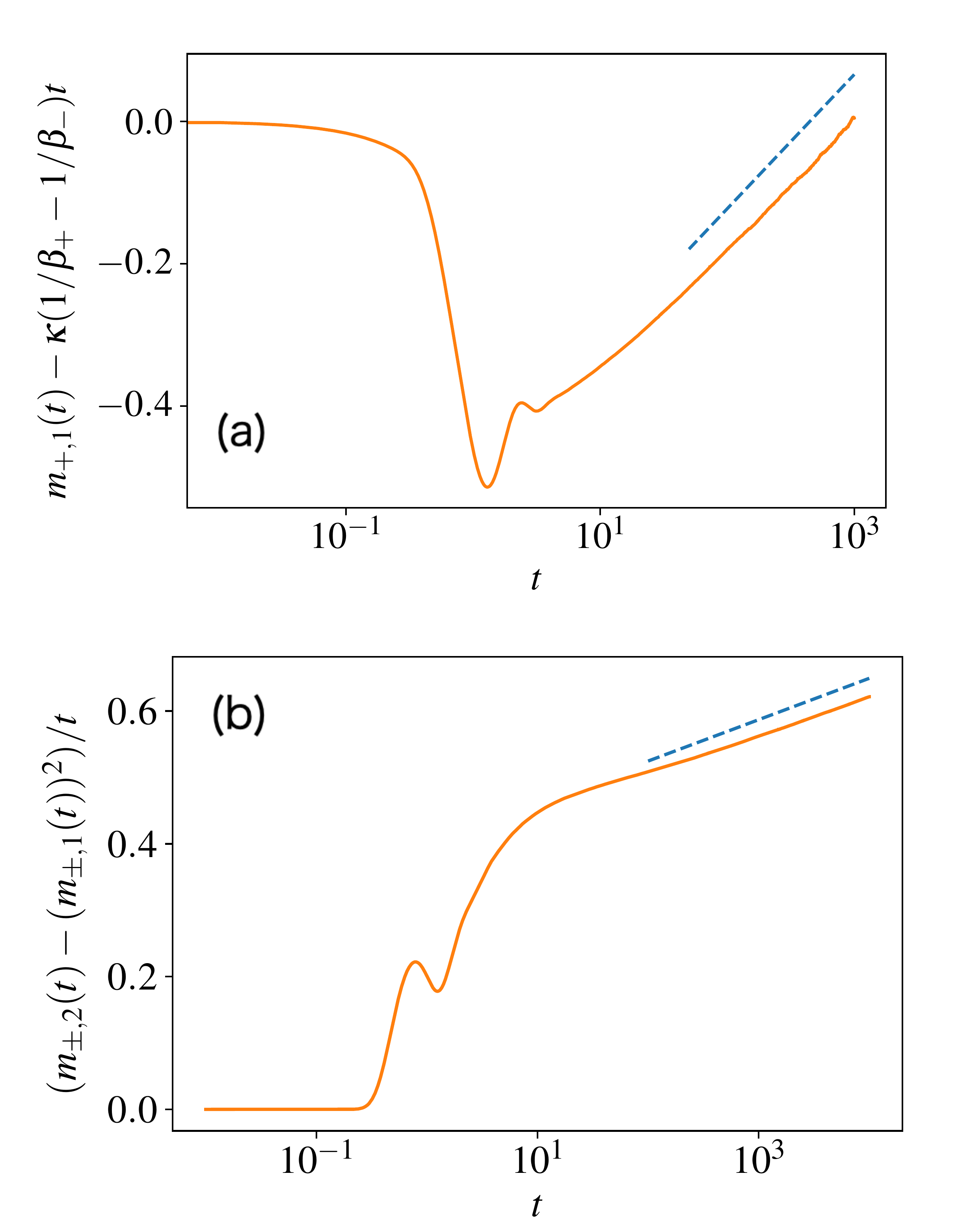}
	 \caption{ $m_{+,1}(t)-\kappa (1/ \beta_{+} - 1/ \beta_{-})$ (a) and $m_{\pm,2}(t) / t^2-(m_{\pm,1}(t))^2/t^2$ (b) obtained from numerical simulations (with $10^8$ samples) are plotted as orange lines. $\beta_+=1$, $\beta_-=2$. Blue dashed lines are $(\kappa^2 / 2) (\beta_+ + \beta_-)    \left( 1 / \beta_+ - 1 / \beta_- \right )  (\ln t) / t+ \rm const.$ and  $ \kappa^3(\beta_++\beta_-)\left( 1 / \beta_+ - 1 / \beta_- \right)^2 + \rm const.$ The slopes of the numerical-simulation results in semi-log scale converge to those of the dashed reference lines, showing the validity of \eqref{eq:mj} and \eqref{eq:convergencevarcurrent}. }
	\label{fig:particle:current}
	\end{center}
\end{figure}

We numerically simulate the one-particle model to check the validity of \eqref{eq:mj} and \eqref{eq:convergencevarcurrent}. We estimate $m_{+,1}(t)$ and $m_{+,2}(t)$ from the numerical simulations, and plot $m_{+,1}(t)-\kappa (1/ \beta_{+} - 1/ \beta_{-})$ and $m_{\pm,2}(t) / t^2-(m_{\pm,1}(t))^2/t^2$ in Fig.~\ref{fig:particle:current} (a,b). In the same figures, we also plot 
$(\kappa^2 / 2) (\beta_+ + \beta_-)    \left( 1 / \beta_+ - 1 / \beta_- \right )  (\ln t) / t+ \rm const.$ and  $ \kappa^3(\beta_++\beta_-)\left( 1 / \beta_+ - 1 / \beta_- \right)^2 + \rm const.$ as blue dashed lines. We observe that the slopes of the orange lines in semi-log scale asymptotically converge to those of blue dashed lines. This demonstrates \eqref{eq:mj} and \eqref{eq:convergencevarcurrent}.

\section{Discussion}
\label{sec:discussion}

\subsection{A counting process with smaller power-law exponents}

In the first part of this article, we studied a counting process $N_t$ with two heavy-tail waiting time distributions: the Pareto distribution with $\alpha=3$ and the inverse Rayleigh distribution. These two waiting time distributions have an asymptotic form $1/\tau^3$ when the waiting time $\tau$ is large, implying that the variance of the waiting time $\mathbb E[\tau^2]$ diverges.  Because of this divergence, we discussed that the scaled variance $c_2(t) t$ of the counting process $N_t$ also diverges in the large $t$ limit. We indeed derived that it is asymptotically proportional with $\ln(t)$, diverging as $t\rightarrow \infty$.

A natural question would be, can we get a similar result with a waiting time distribution that has an asymptotic form $1/\tau^\alpha$ with $\alpha>3$? As demonstrated in Appendix~\ref{sec:appendix_kthmoment}, one can formulate a general framework, for the Pareto distribution, to derive analytical expressions of the Laplace transform of $\mathbb E[N_t^k]$ $(k=1,2,3,...)$ for any $\alpha$. As an example, we computed the first, second and  third moments for $\alpha=4$, from which we show the third cumulant of $N_t/t$ has an asymptotic form $ \ln(t)/t^2$ when $t$ is large. This indicates that the  third-order cumulant multiplied by $t^2$ is asymptotically proportional with $\ln(t)$, which is also diverging in the large $t$ limit.

For the counting process with a general fat tail waiting time distribution (that has a power law decay as $t\rightarrow \infty$), an existence of the affine part in the scaled cumulant generating function (sCGF) $G(s)=\lim_{t \rightarrow \infty}(1/t) \ln \mathbb E[e^{s N_t}]$ has been proven \cite{horii2022large}. When the sCGF is analytic, it can be expanded using scaled cumulants $\bar c_i$ ($i=1,2,...$) as $G(s) =  \sum_{i=1}^{\infty} (\bar c_i / i ! ) s^i$ by definition, where $\bar c_i$ is defined as $\lim_{t\rightarrow \infty} c_i  t^{i-1}$ with the $i$-th order cumulant $c_i$ of $N_t/t$. In the presence of the affine part, sCGF is not analytic, implying that some scaled cumulants $\lim_{t\rightarrow \infty} c_i  t^{i-1}$ diverge. Based on the observation above, we conjecture that  the $k$-th order scaled cumulant converges when $k<\alpha-2$. When $k=\alpha-2$, the $k$-th order cumulant $c_i$ increases proportionally with $\ln(t) / t^{i-1}$, resulting in $\ln(t)$ divergence of $\lim_{t\rightarrow \infty} c_i t^{i-1}$. It is an interesting future work to study this conjecture.

\subsection{Many particles confined in the two hot walls}

In the second part of this article, we studied a particle confined in the two walls in different temperatures, and observed that the scaled variance diverges proportionally with $\ln(t)$. Here we discuss if we can observe the same divergence in many-body particles confined in the walls.

\begin{figure*}[htbp]
\centering
\includegraphics[clip,width=0.99\linewidth]{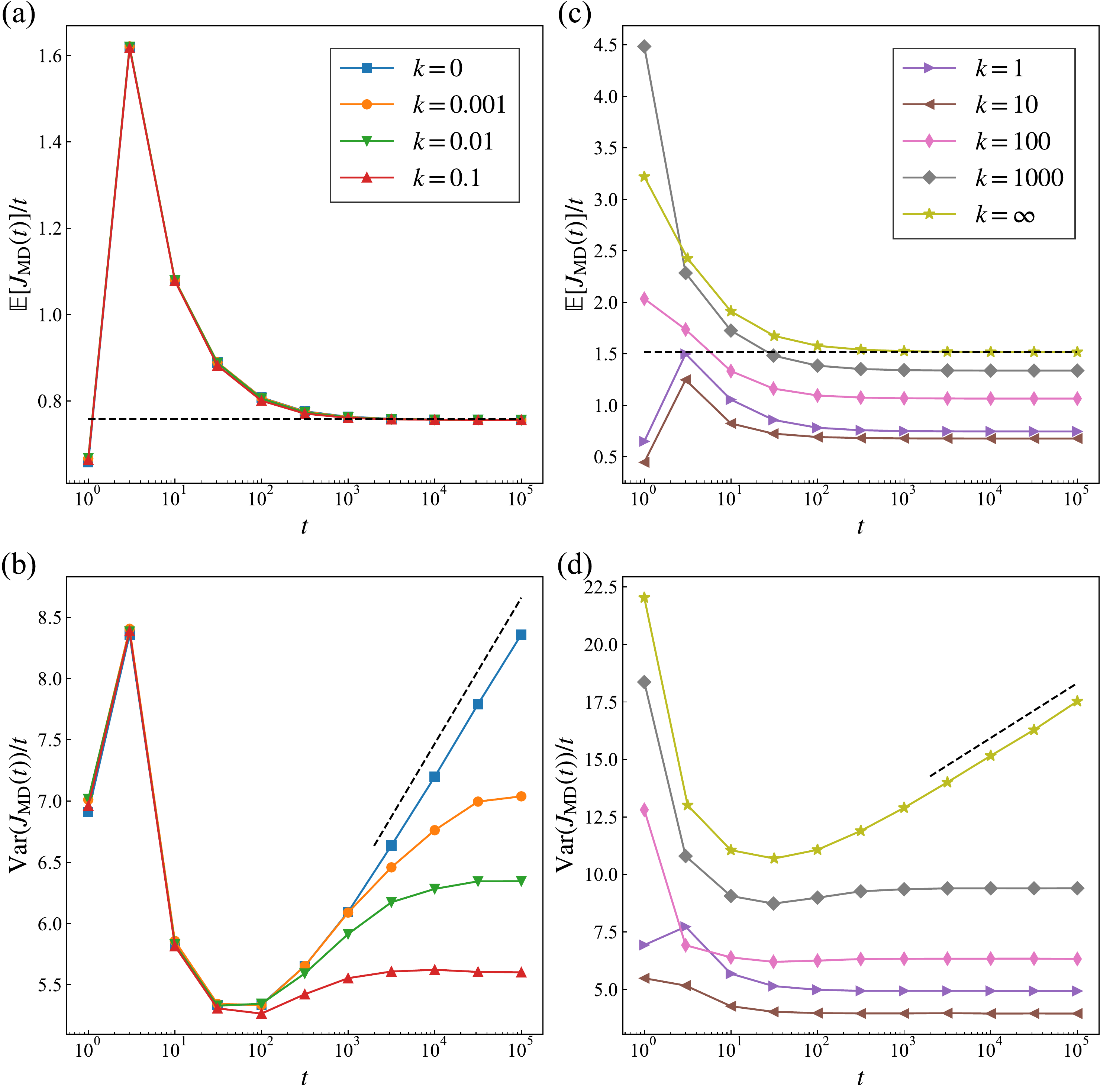}
\caption{Statistical properties of $J_{\mathrm{MD}}(t)$ over time for different particle softnesses averaged over $10^6$ samples when $\mathcal{N}=3$, $D=1$, $L=5$, $\beta_{+}=1/3$, and $\beta_{-}=1$. (a) $\mathbb{E}[J_{\mathrm{MD}}(t)]/t$ versus $t$ for $k=0$, $0.001$, $0.01$, and $0.1$. The dashed line is $\mathbb{E}[J_{\mathrm{MD}}(t)]/t=\mathcal{N}\kappa(\beta_{+}^{-1}-\beta_{-}^{-1})/(L-D)$. (b) $\mathrm{Var}(J_{\mathrm{MD}}(t))/t$ versus $t$ for $k=0$, $0.001$, $0.01$, and $0.1$. The dashed line is $\mathrm{Var}(J_{\mathrm{MD}}(t))/t=\mathcal{N}\kappa^{3}(\beta_{+}+\beta_{-})(\beta_{+}^{-1}-\beta_{-}^{-1})^{2}\ln(t)/(L-D)+\mathrm{const}$. (c) $\mathbb{E}[J_{\mathrm{MD}}(t)]/t$ versus $t$ for $k=1$, $10$, $100$, $1000$, and $\infty$. The dashed line is $\mathbb{E}[J_{\mathrm{MD}}(t)]/t=\mathcal{N}\kappa(\beta_{+}^{-1}-\beta_{-}^{-1})/(L-\mathcal{N}D)$. (d) $\mathrm{Var}(J_{\mathrm{MD}}(t))/t$ versus $t$ for $k=1$, $10$, $100$, $1000$, and $\infty$. The dashed line is $\mathrm{Var}(J_{\mathrm{MD}}(t))/t=\mathcal{N}\kappa^{3}(\beta_{+}+\beta_{-})(\beta_{+}^{-1}-\beta_{-}^{-1})^{2}\ln(t)/(L-\mathcal{N}D)+\mathrm{const}$.}
\label{fig:MD}
\end{figure*}

One-dimensional hard-core interacting particles exchange their velocities when they collide. The dynamics of these particles can thus be exactly mapped to the dynamics of non-interacting many-body particles. 
Let $J^{\mathcal{N},L,D}_{\infty}(t)$ and $J^{\mathcal{N},L,D}_{0}(t)$ be the energy currents of $\mathcal{N}$ hard-core interacting and non-interacting particles of diameter $D$ confined in a one-dimensional box of size $L$, respectively.
Then, we get
\begin{align}
 \mathbb{E}\left[ \frac{J^{\mathcal{N},L+\mathcal{N}D,D}_{\infty}(t)}{t}\right] &= \mathbb{E}\left[ \frac{J^{\mathcal{N},L+D,D}_{0}(t)}{t}\right] 
 \\
 &= \mathcal{N}\frac{m^{L}_{\pm,1}(t)}{t},
 \\
 \mathrm{Var}\left( \frac{J^{\mathcal{N},L+\mathcal{N}D,D}_{\infty}(t)}{t}\right) &= \mathrm{Var}\left( \frac{J^{\mathcal{N},L+D,D}_{0}(t)}{t}\right)
 \\
 &=\mathcal{N}\mathrm{Var}\left( \frac{J^{L}_{\pm}(t)}{t}\right),
\end{align}
where $m^{L}_{\pm,1}(t)$ and $\mathrm{Var}(J^{L}_{\pm}(t)/t)$ are given in Eqs.~(\ref{eq:mjL}) and (\ref{eq:convergencevarcurrentL}), respectively.
This implies that the logarithmic divergence of the scaled variance should be observed in hard-core interacting systems. 
In soft-core interacting systems, on the other hand, the same mapping cannot be used. This is because of the collisions involving more than two particles, where the exchange rule of velocities no longer holds. To demonstrate this insight,  we have performed simulations of hard-core and soft-core interacting particles. The details of the simulations are explained in Appendix \ref{sec:MDdetail}, and the results are shown in Fig.~\ref{fig:MD}, where $J_{\mathrm{MD}}(t)$ is the total energy transferred to the colder wall from time $0$ to $t$, and $k$ is a parameter corresponding to the softness of particles. Note that $k=\infty$ corresponds to the case of the hard-core interacting system. We observed that the  $\ln(t)$ divergence disappears as soon as particles start to interact via soft-core interactions. It is an interesting future problem to develop a framework to quantitatively understand the disappearance of the divergence in soft-core particles. 

\subsection{Related studies}

Finally, we list related studies. Studying a variance in a process that is defined with power-law decaying distribution is not something new. In \cite{metzler2014anomalous}, several anomalous diffusion models were studied using continuous-time random walk, and revealed anomalous scaling in their diffusion coefficients. These anomalous scalings were argued to be universally observed in transports in random media \cite{barkai2020packets}. One of the authors also studied a single big jump principle, which states that the sum of random variables can be approximated by their maximum when the probability distribution of the variables has a power-law \cite{b6}.

Singularities of large deviation functions of time-cumulative quantities are also known as dynamical phase transitions, and have been studied in many physical models, such as glass formers \cite{hedges2009dynamic, garrahan2009first, jack2010large, pitard2011dynamic, limmer2014theory, nemoto2017finite, speck2012first}, lattice gas models \cite{bodineau2005distribution, appert2008universal, bodineau2008long, bodineau2012finite, baek2017dynamical, shpielberg2017geometrical, shpielberg2018universality},  diffusive hydrodynamic equations \cite{bertini2005current, hurtado2011spontaneous, tizon2017structure}, and high-dimensional chaotic dynamics \cite{tailleur2007probing, laffargue2013large, bouchet2014stochastic}  and active matters \cite{vaikuntanathan2014dynamic, cagnetta2017large, whitelam2018phase, nemoto2019optimizing}. Finite-size scalings of the large deviation functions have been performed in several works (see an interesting recent work \cite{PhysRevLett.128.090605} for example), but variance scalings have not been intensively studied in this field yet.

\begin{acknowledgments}
The authors thank S. Sasa for useful comments. H.H. is in the Cofund MathInParis PhD project that has received funding from the European Union's Horizon 2020 research and innovation programme under the Marie Sk lodowska-Curie grant agreement No.~754362~\includegraphics[width=0.7cm]{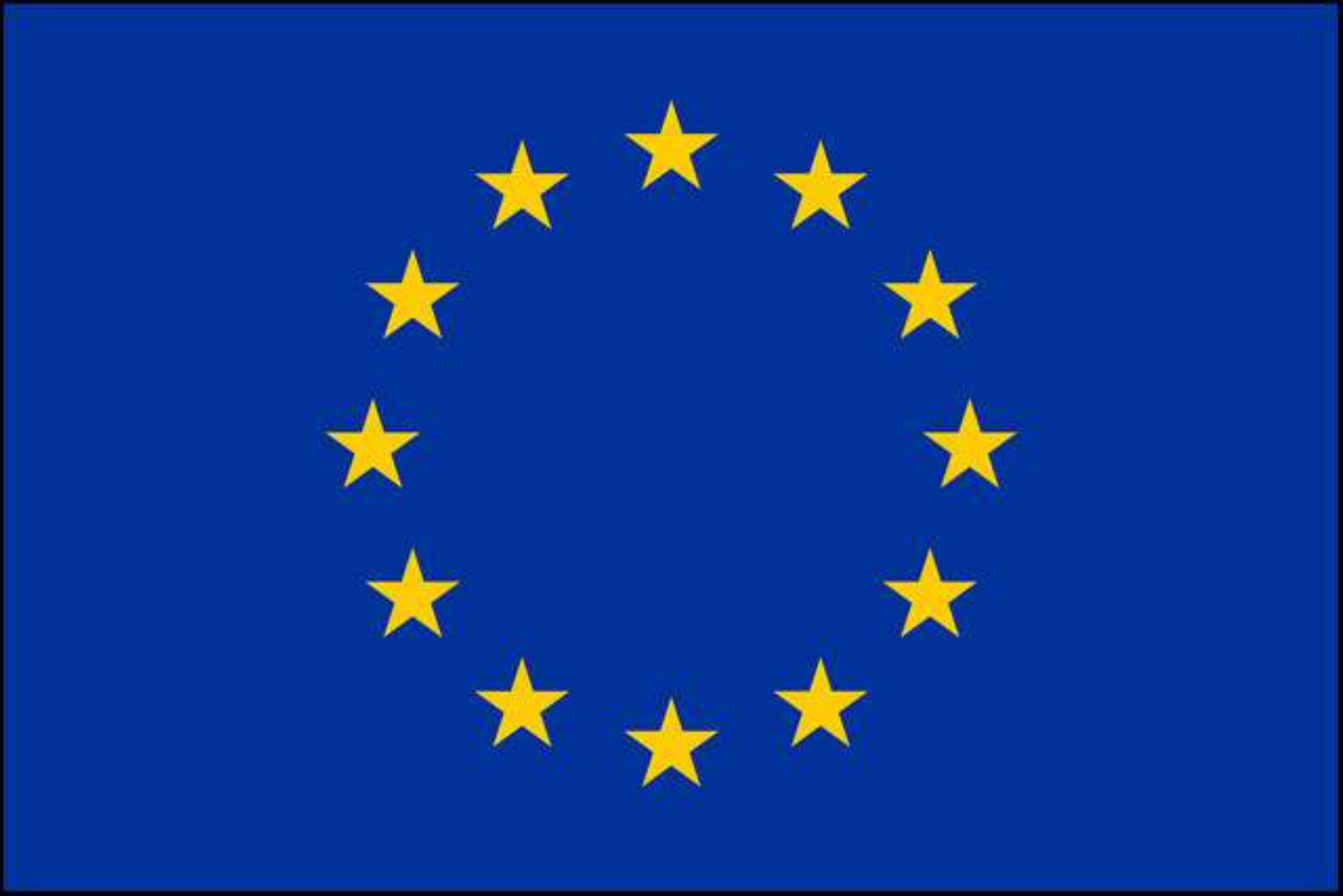}. R.L. is supported by the ANR-15-CE40-0020-01
grant LSD.
This work was supported by JSPS KAKENHI under Grants No.~JP17H01148 and No.~JP20J00003.
\end{acknowledgments}

\appendix

\section{$k$-th moment of a counting process with heavy-tailed distributions}
\label{sec:appendix_kthmoment}

Here, we derive the $k$-th moment of a counting process $N_{t}$ (with a waiting time density $p$) by using a renewal equation.
The moment-generating function $M_{h}(t)$ is defined by
\begin{align}
 M_{h}(t) \equiv \mathbb{E}\left[ e^{hN_{t}}\right].
\end{align}
Using
\begin{align}
 \mathbb{E}\left[ e^{hN_{t}}\right] &= \int_{0}^{\infty}\mathrm{d}u\; \mathbb{E}\left[ e^{hN_{t}}\vert \tau_{1}=u\right] p(u)
 \notag\\
 &= \int_{0}^{t}\mathrm{d}u\; \mathbb{E}\left[ e^{h(N_{t-u}+1)}\right] p(u) 
 \notag\\
 &\qquad + \int_{t}^{\infty}\mathrm{d}u\; p(u),
\end{align}
we obtain the following renewal equation
\begin{align}
 M_{h}(t) =  e^{h}\int_{0}^{t}\mathrm{d}u\; M_{h}(t-u) p(u) + \int_{t}^{\infty}\mathrm{d}u\; p(u).
\end{align}
The Laplace transform of this equation gives
\begin{align}
 \tilde{M}_{h}(s) = e^{h}\tilde{M}_{h}(s) \tilde{p}(s) + \frac{1-\tilde{p}(s)}{s},
\end{align}
which leads to
\begin{align}
 \tilde{M}_{h}(s) = \frac{1}{s}\frac{1-\tilde{p}(s)}{1-\tilde{p}(s)e^{h}}.
\label{eq:L_MGF}
\end{align}
Using
\begin{align}
 \tilde{m}_{1}(s) = \frac{\tilde{p}(s)}{s(1-\tilde{p}(s))},
\end{align}
we can rewrite (\ref{eq:L_MGF}) as
\begin{align}
 \tilde{M}_{h}(s) &= \frac{1}{s}\frac{1}{1-s\tilde{m}_{1}(s)(e^{h}-1)} 
 \notag\\
 &= \sum_{q=0}^{\infty}(e^{h}-1)^{q}s^{q-1}\left[ \tilde{m}_{1}(s)\right]^{q}.
\end{align}
Because
\begin{align}
 \lim_{h\to 0}\frac{\mathrm{d}^{k}}{\mathrm{d}h^{k}}(e^{h}-1)^{q} &= \sum_{i=0}^{q}\begin{pmatrix}q\\i\end{pmatrix}i^{k}(-1)^{q-i},
 \\
 \lim_{h\to 0}\frac{\mathrm{d}^{k}}{\mathrm{d}h^{k}}(e^{h}-1)^{q} &= 0, \quad \text{for}\quad k < q,
\end{align}
and
\begin{align}
 \lim_{h\to 0}\frac{\mathrm{d}^{k}}{\mathrm{d}h^{k}}\tilde{M}_{h}(s) = \tilde{m}_{k}(s),
\end{align}
we have
\begin{align}
\label{eq:app:generalmk}
 \tilde{m}_{k}(s) = \sum_{q=1}^{k} \left[ \sum_{i=1}^{q}\begin{pmatrix}q\\i\end{pmatrix}i^{k}(-1)^{q-i}\right] s^{q-1}\left[ \tilde{m}_{1}(s)\right]^{q}.
\end{align}

Let us now consider the Pareto distribution \eqref{eq:Pareto_countingprocess} with $\alpha=4$ as the waiting time density. In this case, we have
\begin{align}
 \tilde{m}_1(s)&= \frac{2}{s^2}+\frac{1}{s}+o\left(\frac{1}{s}\right),
 \\
 \tilde{m}_2(s)&= \frac{8}{s^3}+\frac{10}{s^2}+\frac{16\ln(s)}{s}+o\left(\frac{\ln(s)}{s}\right),
 \\
 \tilde{m}_3(s)&= \tilde{m}_1(s)+6s\tilde{m}^2_1(s)+6s^2\tilde{m}^3_1(s)
 \notag\\
 &=\frac{48}{s^4}+\frac{96}{s^3}+\frac{144\ln(s)}{s^2}+o\left(\frac{\ln(s)}{s^2}\right)
\end{align}
from \eqref{eq:app:generalmk} for large $s$.
Using the following inverse Laplace transform
\begin{equation}
	\int^{\infty}_0 e^{-st}\ln(t) dt =\left(-\frac{\ln(s)+\gamma}{s}\right),
\end{equation}
we calculate the inverse Laplace transform of $\tilde{m}_k(s)$ as
\begin{eqnarray}
	m_1(t)&\sim&2t+1,\\
	m_2(t)&\sim&4t^2+10t-16\ln(t),\\
	m_3(t)&\sim&8t^3+48t^2+206t-144t\ln(t)
\end{eqnarray}
as $t\rightarrow \infty$. The  second cumulants $c_2$ defined as \eqref{defc2} and the third cumulant  $c_3$ (defined as the third cumulant of $N_t/t^3$) are then given by
\begin{align}
 c_2(t)&=\frac{6}{t}+o\left(\frac{1}{t}\right),
 \\
 c_3(t)&=\frac{m_3(t)-3m_1(t)m_2(t)+2m_1^3(t)}{t^3}
 \notag\\
 &=-\frac{48\ln (t)}{t^{2}}+o\left( \frac{\ln (t)}{t^{2}}\right).
\end{align}

\section{Tauberian theorem}
\label{sec:tauberian}

The Tauberian theorem is stated in \cite{feller2008introduction} Ch.XIII.5, theorem 4.  In our context it can be stated as follows.  If the Laplace transform $\tilde m$ of the renewal function $m$ satisfies
$$
\tilde m(s)\sim \frac{1}{s^{\rho}}L\left(\frac 1 s\right), \quad s\to 0
$$
for some $\rho>0$ and some slowly varying ({\it i.e.,} a function $L$ is slowly varying if for any $x>0$, $\lim_{t\to\infty} \frac{L(xt)}{L(t)}=1$) function $L$ then
\begin{equation}
m(t)\sim \frac{1}{\Gamma(\rho)} t^{\rho-1}L(t).
\label{tauberian}
\end{equation}

\section{Simulation detail} \label{sec:MDdetail}

$\mathcal{N}$ particles of mass $m$ and diameter $D$ are lined up on a line $[0,L]$.
Let $(r_{i},p_{i})$ be the position and momentum of the $i$th particle.
The total energy transferred to the right wall from time $0$ to $t$ is defined by
\begin{align}
 J_{\mathrm{MD}}(t) = \sum_{i}\sum_{k_{i}}\left\{ \frac{\left\vert p_{i}(t_{k_{i}}-0)\right\vert^{2}}{2m}-\frac{\left\vert p_{i}(t_{k_{i}}+0)\right\vert^{2}}{2m}\right\}
\end{align}
with $0\leq t_{k_{i}}\leq t$, where $t_{k_{i}}\pm 0$ is the time just before/after the $i$th particle collides with the right wall for the $k_{i}$th time.

For the case of the soft-core interacting system, a short-range interaction potential $\Phi$ between two particles is given by
\begin{align}
 \Phi (\vert r_{i}-r_{j}\vert) = \frac{k}{2}\left( D-\vert r_{i}-r_{j}\vert\right)^{2}\Theta(D-\vert r_{i}-r_{j}\vert),
\end{align}
where $\Theta$ is the Heaviside step function, and $k$ is a parameter corresponding to the softness of particles.
The boundary condition is the same as explained in Sec.~\ref{subsec:Model}.
Using the second-order symplectic integrator, we numerically solved the equations of motion for the particles, and calculated $\mathbb{E}[J_{\mathrm{MD}}(t)]$ and $\mathrm{Var}(J_{\mathrm{MD}}(t))$ for various values of $k$.
In the simulation, we set the parameter values as $\mathcal{N}=3$, $L=5$, $m=D=1$, $\beta_+=1/3$, and $\beta_-=1$.
The time-discretization step-size was set to $0.01$.

For the case of the hard-core interacting system (denoted by $k=\infty$), we performed event-driven simulations in which two particles instantaneously exchange velocities when they come into contact.
The boundary condition and the parameter values were the same as for the soft-core particle system.

\end{document}